\documentclass[a4paper,11pt]{article}
\usepackage{pos}
\usepackage{cleveref}

\definecolor{hu-berlin-blue}{RGB}{0,65,137} % HEX 004189

\title{In-medium static quark potential from spectral functions on realistic HISQ ensembles}
\author*[b]{Gaurang Parkar}
\author{Dibyendu Bala$^a$, Olaf Kaczmarek$^a$, Rasmus Larsen$^b$, Swagato Mukherjee$^c$, Peter Petreczky$^c$, Alexander Rothkopf$^b$,
Johannes Heinrich Weber$^d$
}

\affiliation{
$^a$ Fakult\"at f\"ur Physik, Universit\"at Bielefeld, D-33615 Bielefeld, Germany\\
$^b$ Faculty of Science and Technology, University of Stavanger, NO-4036 Stavanger, Norway,\\
$^c$ Physics Department, Brookhaven National Laboratory, Upton, New York 11973, USA\\
$^d$Institut f\"ur Physik \& IRIS Adlershof, Humboldt-Universit\"at zu Berlin, D-12489 Berlin, Germany
}

\emailAdd{gparkar@ux.uis.no}

\abstract{
We explore the interactions between a quark anti-quark pair in a thermal medium based on lattice QCD ensembles with $N_f = 2+1$ dynamical HISQ flavors. Our dataset spans the phenomenologically relevant temperature range between T=140MeV-2GeV based on lattice sizes 
$N_\tau=10,12$ and $16$, with an aspect ratio of $N_\sigma/N_\tau=4$.
The peak position $\Omega$ and the width $\Gamma$ of the  spectral function of Wilson-line correlators in Coulomb gauge is computed. We assess the information content in the correlation functions and deploy three complementary strategies to reconstruct spectral information: model fits, Pad\'e approximation and the Bayesian BR method. Limitations of each approach are carefully assessed.}

\FullConference{%
  The 38th International Symposium on Lattice Field Theory, LATTICE2021 26th-30th July, 2021
Zoom/Gather@Massachusetts Institute of Technology
}

\begin{document}
\maketitle

\section{Introduction}
A first principles understanding of bound states of a heavy quark and anti-quark pair (quarkonia) hold the key to probing the existence and properties of the quark gluon plasma (QGP) in heavy-ion collisions (HIC). Lattice QCD has made vital contributions to estimating the properties of quarkonium at zero temperature and to the thermodynamic properties of the QGP. Static quarks at finite temperature can be studied via the Wilson loop or Wilson line correlator in Coulomb gauge, which itself can be related to the evolution of quarkonia via effective field theories.

The separation of scales $M \gg Mv \gg Mv^2, M \gg \Lambda_{QCD}$ (where $M$ is the heavy quark mass and $v$ is the relative velocity in the bound state) tells us that if we consider processes at a scale below $M$ a non-relativistic description (NRQCD) of quarks is possible. If we set the energy cutoff to $Mv$ and integrate out degrees of freedom that cannot be excited above $Mv^2$, a different effective field theory called potential NRQCD (pNRQCD) ensues. The term potential in pNRQCD refers to the non-local Wilson coefficients in the Lagrangian, the lowest order of which can be related to the Wilson loop and its spectrum.

At zero temperature the evolution of static quark anti-quark pairs after time coarse-graining can be described by a Schr\"odinger-like equation for the Wilson loop with a non-relativistic potential $V(r)$ \cite{Bali:2000gf}
\begin{equation}
\label{potential}
i \partial_t W_\Box (t,r) = \Phi(t,r)W_\Box(t,r), \quad
V(r) = \lim_{t \rightarrow \infty} \Phi(t,r).
\end{equation}

Our long-term goal lies in ascertaining whether the potential picture also holds at T>0. And if it does, how the form of the potential changes from the zero temperature case.

The real-time evolution of the Wilson loop referred to above can be studied on the lattice via its spectral function $\rho_\square$, which acts as a link between the real-time $W_{\Box}(r,t)$ and imaginary-time Wilson Loop $W_{\Box}(r,\tau)$ \cite{Rothkopf:2011db}
\begin{equation}
\label{spectral_link}
 W_\Box (r,t) = \int d \omega e^{-i \omega t}\rho_\Box (r,\omega)  \leftrightarrow
\int d \omega e^{-\omega \tau} \rho_\Box (r,\omega) = W_\Box (r,\tau).
\end{equation}
In order to extract the spectral function from the Wilson loop $W_{\Box}(t,r)$ we must invert \cref{spectral_link}. However, in practice the presence of noisy data and the availability of only a few data points along imaginary time make the inversion an ill-posed problem. Hard thermal loop perturbative computations (HTL) show that there exists a dominant peak structure in the spectral function related to a complex potential with a screened real- \cite{Laine:2006ns} and finite imaginary part. 

A first step toward establishing whether \cref{potential} holds requires us to investigate the spectral structure of the Wilson loop, in particular we are interested in the position $\Omega$ and width $\Gamma$ of the lowest lying spectral structure, which has been related to its late time evolution in \cite{Burnier:2012az}.

\section{Lattice Setup}
\label{setup}
We performed calculations on Wilson loop and Wilson line correlators from (2+1)-flavour QCD configurations generated by HotQCD and TUMQCD collaborations \cite{Bazavov:2018wmo,Bazavov:2017dsy,HotQCD:2014kol,Bazavov:2019qoo}. 
The Highly Improved Staggered Quark (HISQ) action was used to generate  ensembles providing us with $2-6 \times 10^4$ gauge configurations. $N_\sigma^3 \times N_\tau$ lattices were used with $N_\tau = 10,12,16$ to control lattice spacing effects and the aspect ratio $N_\sigma/N_\tau$ was set to 4 to control finite volume effects. The Wilson Line correlators were calculated in Coulomb Gauge. The fixed-box approach was used to scan temperature range from 140Mev to 2GeV, with $T=0$ simulations available for scale setting. The ratio $m_l/m_s$ was set to 1/20 for low temperatures (T<300 MeV) and 1/5 for some ensembles at higher 
%high 
temperatures (T>300 MeV).

\section{Cumulants and HTL comparison} 
To analyse properties of lattice data we define cumulants of the correlation function.
\begin{eqnarray}
m_1(r,\tau,T)=-\partial_{\tau} \ln W(r,\tau,T),\quad m_n=\partial_{\tau} m_{n-1}(r,\tau,T), n>1,
\end{eqnarray}
where the first cumulant $m_1$ is nothing but the effective mass. We find that our data only allows us to extract information up to the first three
%two 
cumulants, due to loss of signal for higher orders.

\begin{figure}[h]
\includegraphics[scale=0.6]{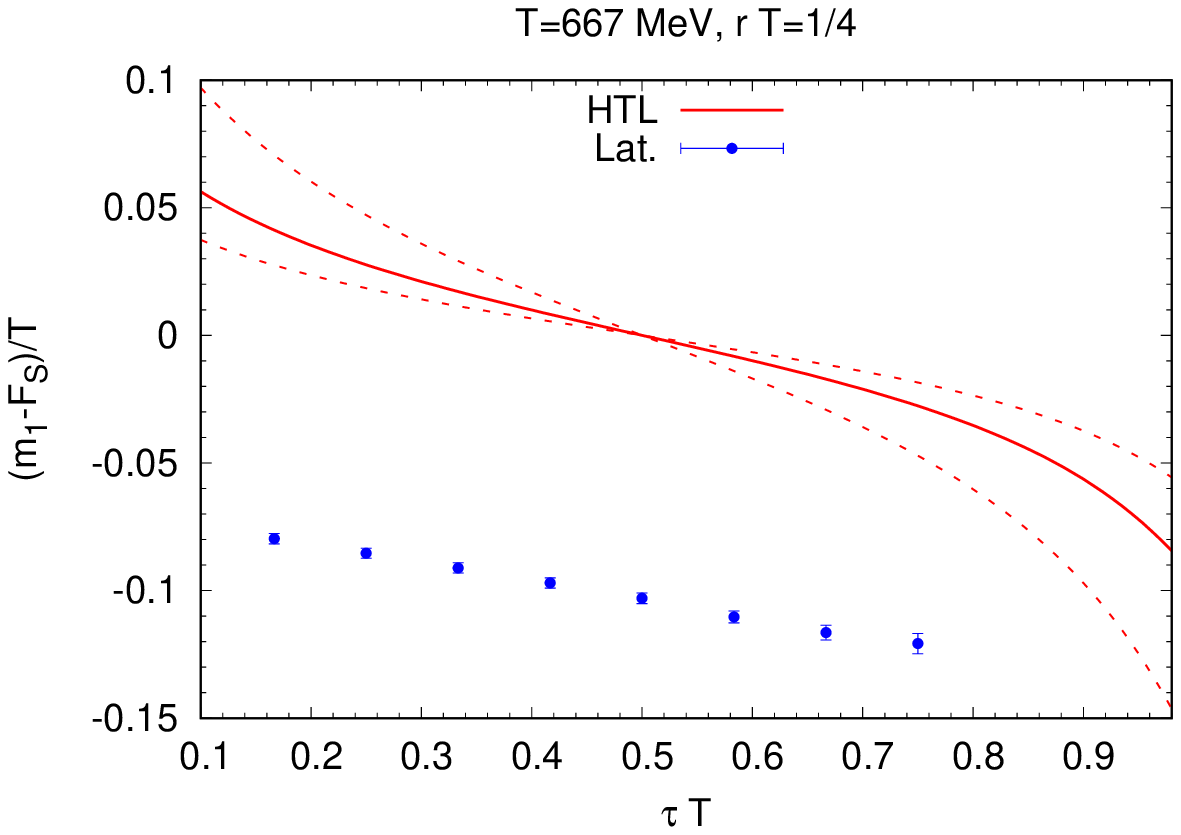}
\includegraphics[scale=0.6]{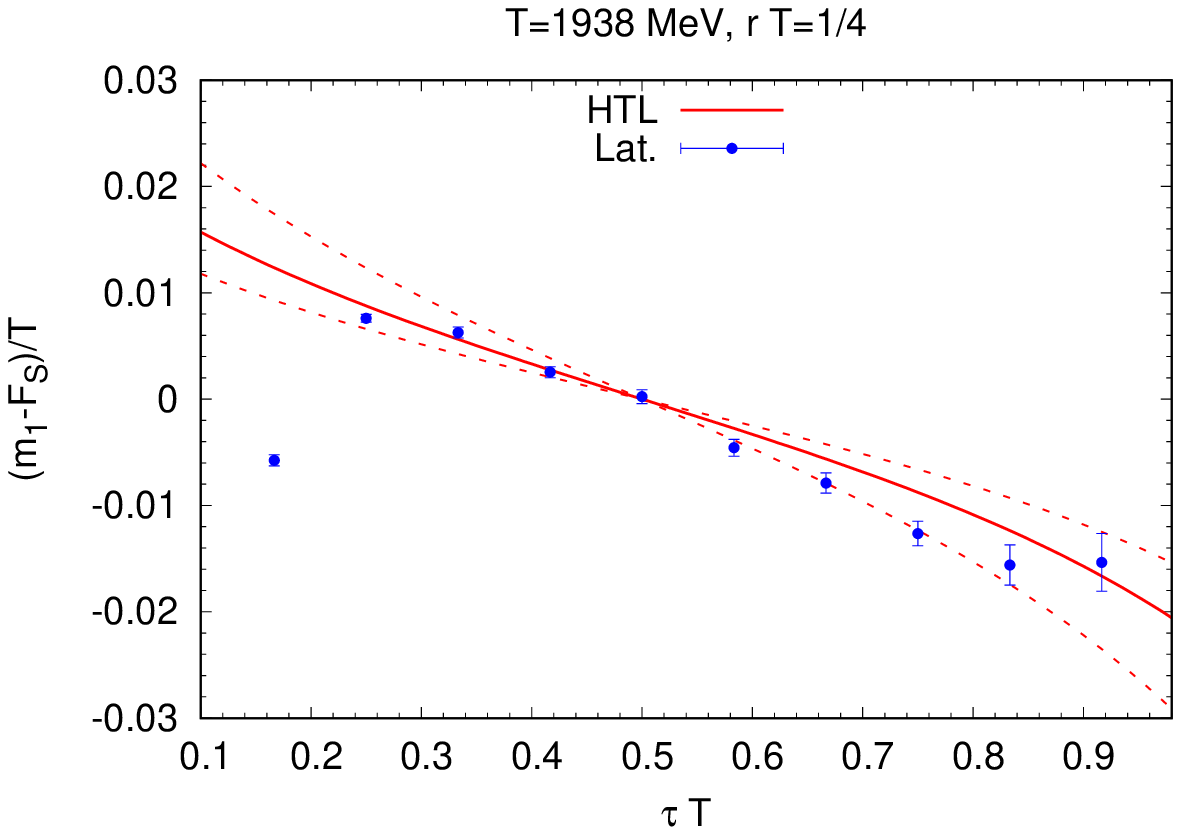}
\caption{Comparison of $m_1-F_S$ calculated on the lattice with HTL results at $T=667$ MeV and $T=1938$ MeV for $rT=1/4$.
The HTL result for $\mu=2 \pi T$ are shown as solid lines. The dashed lines
correspond to variation of the scale $\mu$  by a factor of two.}
\label{cumulant1}
\end{figure}

\begin{figure}[h]
\centering
\includegraphics[scale=.7]{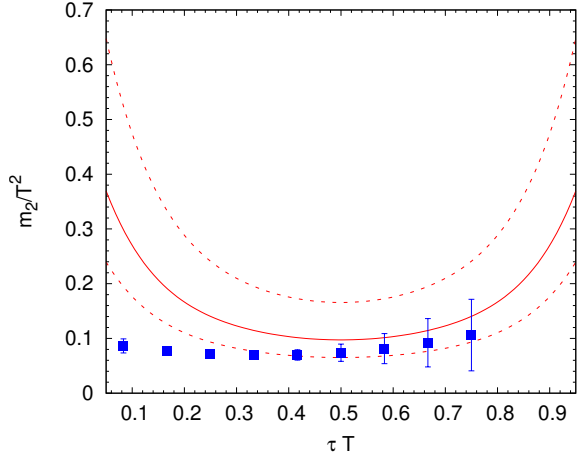}
\caption{The second cumulant of the subtracted Wilson line correlators as function of $\tau$ at $T=667$ MeV calculated
on $N_{\tau}=12$ lattices and in HTL perturbation theory (lines) for
$rT=1/4$. The renormalization scale $\mu$ $\mu=2 \pi T$ are shown as solid lines. The dashed lines
correspond to variation of the scale $\mu$  by a factor of two.}
\label{cumulant2}
\end{figure}

For the analysis we subtract the UV part of the correlator $W$ using T=0 data as described in \cite{Larsen:2019zqv}, since the UV part is temperature independent and assumed to be well separated from the low-lying structures of interest to our study. A comparison of the first cumulant obtained on the lattice with HTL is shown in Fig.\ref{cumulant1}, where we plot the difference of $m_1$ with the color singlet free energies. In HTL this difference is antisymmetric around $\tau=\beta/2$. 
We see that the first cumulant does not agree with HTL results at $T=667$ MeV however, better agreement is observed with increasing temperature e.g. at $T=1938$MeV. A comparison of the second cumulant with HTL at $T=667$MeV is shown in Fig \ref{cumulant2}.
There we again find that the symmetry present in the HTL result is not reproduced by the data. Quantitative agreement however is observed around $\tau = \beta/2$. 

\section{Spectral function using model fits}

As seen in the last section, we can only extract the first two cumulants of the lattice data. While the first 
cumulant
%moment 
carries information on the dominant peak position $\Omega$, the second 
cumulant
%moment 
is related to the width $\Gamma$ of that spectral structure. When we plot the first cumulant based on the subtracted data we see a linear falloff at small $\tau$. This is compatible with a Gaussian peak. Therefore as a first step we model the correlator as
\begin{equation}
\label{gaussian_fit}
\begin{split}
    W(r,\tau,T) = A_P \exp\big[-\Omega(r,T)\tau+\Gamma_G(r,T) ^2 \tau ^2/2\big]+ 
    A_{cut}(r,T) \exp(
    \big[-\omega_{cut}(r,T)\tau\big]
    \end{split}
\end{equation}

\begin{figure}
\centering
\includegraphics[scale=.5]{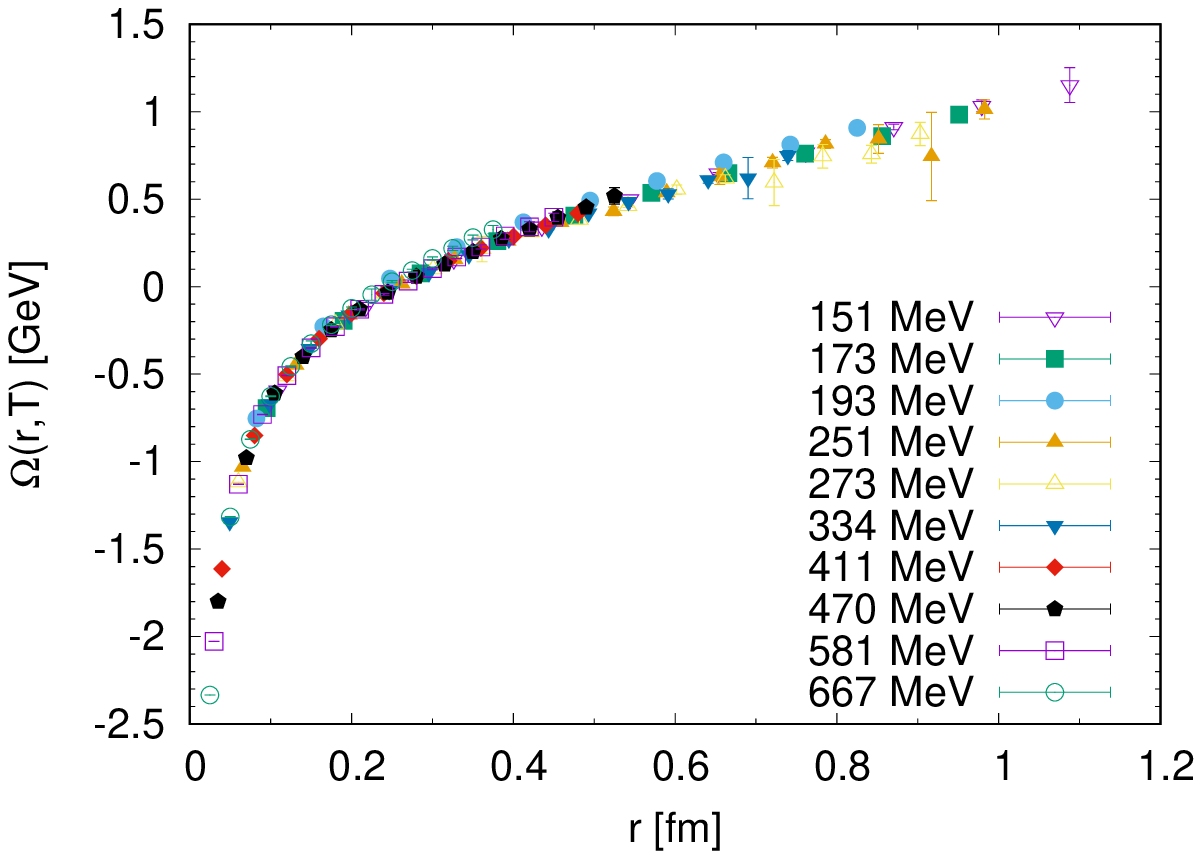}
\includegraphics[scale=.5]{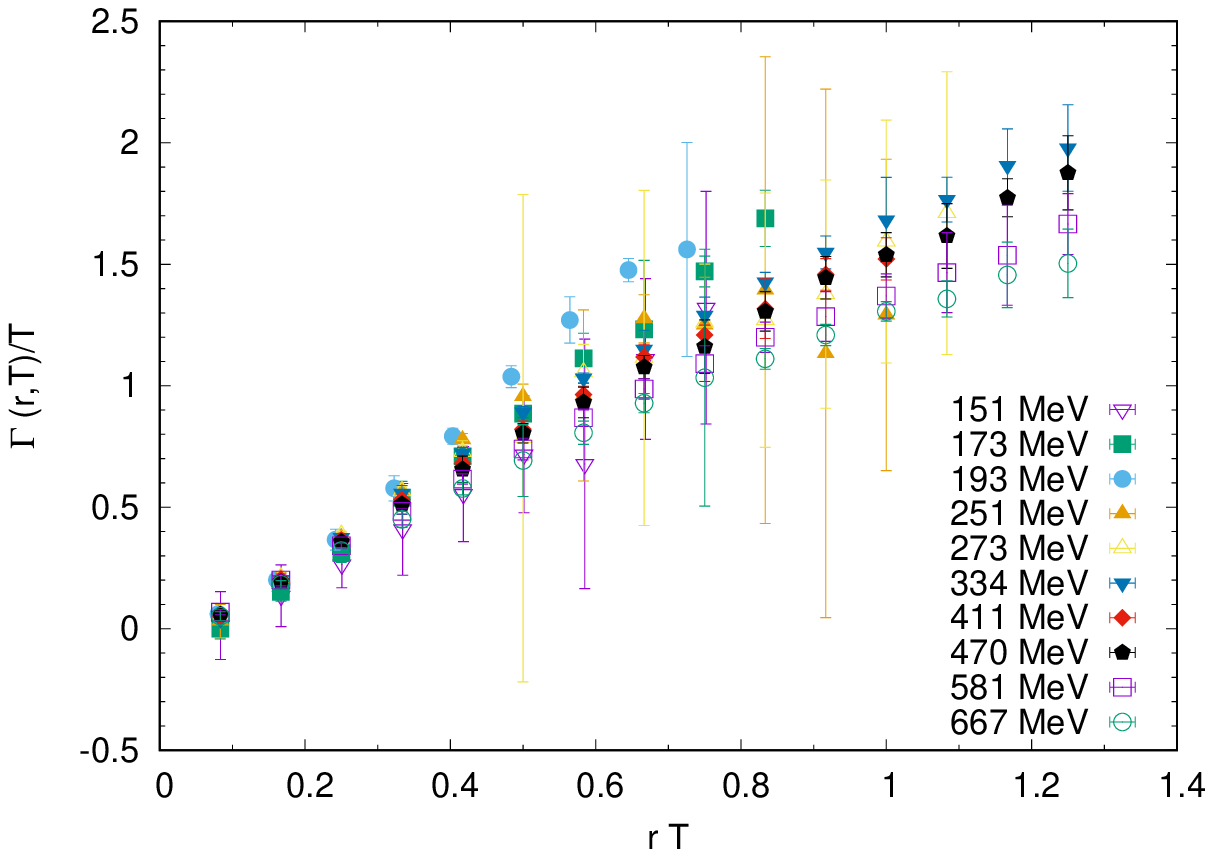}
\caption{The peak position of the spectral function (left figure) and the width  (right figure) as function of the separation $r$ obtained from Gaussian fits of the $N_{\tau}=12$ data.}
\label{Fig:fit_plot}
\end{figure}

Fitting the lattice correlator data with equation \ref{gaussian_fit} we can extract $\Omega$ and 
an effective width $\Gamma=\sqrt{2\ln2}\Gamma^G$ 
%$\Gamma$
as shown in Fig.\ref{Fig:fit_plot}.
We find no significant dependence of $\Omega$ on temperature and the results are similar to the T=0 results. $\Gamma/T$ 
scales as a function of $rT$
%shows a linear dependence on temperature 
within the uncertainties.

\section{Spectral Function Extraction Using Pad\'e}
\label{pade}

The second approach we deploy is the Pad\'e approximation, which on the one hand is model independent but on the other hand very sensitive to noise in the input data. We first start by Fourier transforming the Euclidean correlator into Matsubara frequency space. 
\begin{align}
\displaystyle 
W(r,\tilde \omega_n,T)=\sum_{j=0}^{N_\tau-1}e^{ia \tilde \omega_n j } W(r,j a,T),~\tilde \omega_n=2 \pi n/aN_{\tau}. \label{eq:specdec}
\end{align}
We then project the data onto a basis of rational functions and carry out an interpolation. This step is carried out according to the Schlessinger prescription \cite{PhysRev.167.1411}. Taking an interpolation of data instead of fitting avoids costly minimisation. We analytically continue the rational function by replacing Matsubara frequencies with real-time frequencies. Note that instead naive Fourier frequencies $\hat{\omega_n}$ we use corrected frequencies based on the lattice dispersion relation 
\begin{align}
\tilde \omega_n \rightarrow \omega_n=2 {\rm sin}\big(\frac{\pi n}{ N_\tau}\big)/a.
\end{align}
When we have a rational interpolation function in the Matsubara frequency space we can directly obtain the pole structure by finding roots of the denominator. The pole structure is directly related to the peak position $\Omega$ and width $\Gamma$ of the spectral function. We select the pole that is closest to the real axis to get the dominant peak structure. We identify $\Omega$ with the real part of the dominant pole and $\Gamma$ with the imaginary part.

\begin{figure}[h]
\includegraphics[scale=0.5]{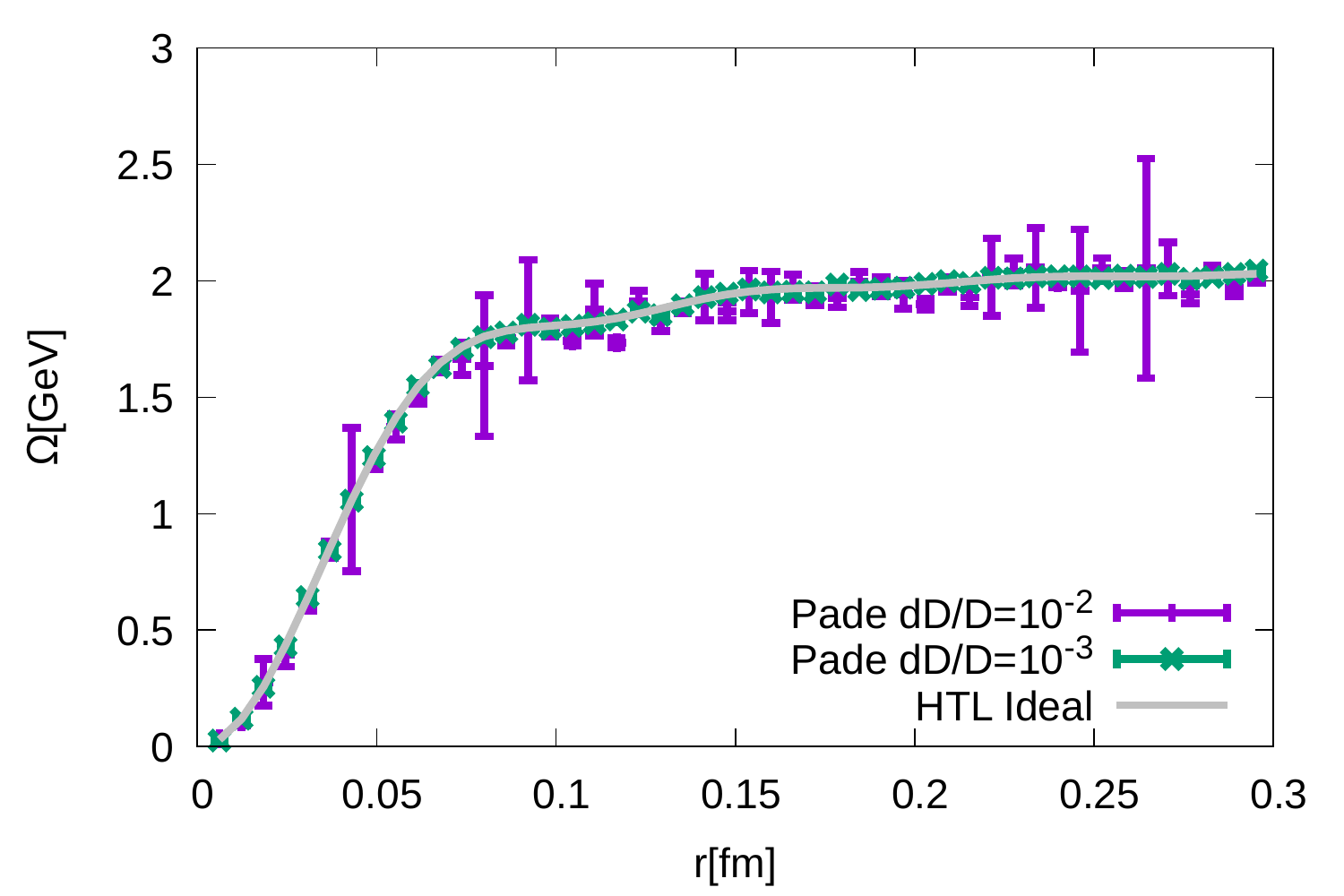}
\includegraphics[scale=0.5]{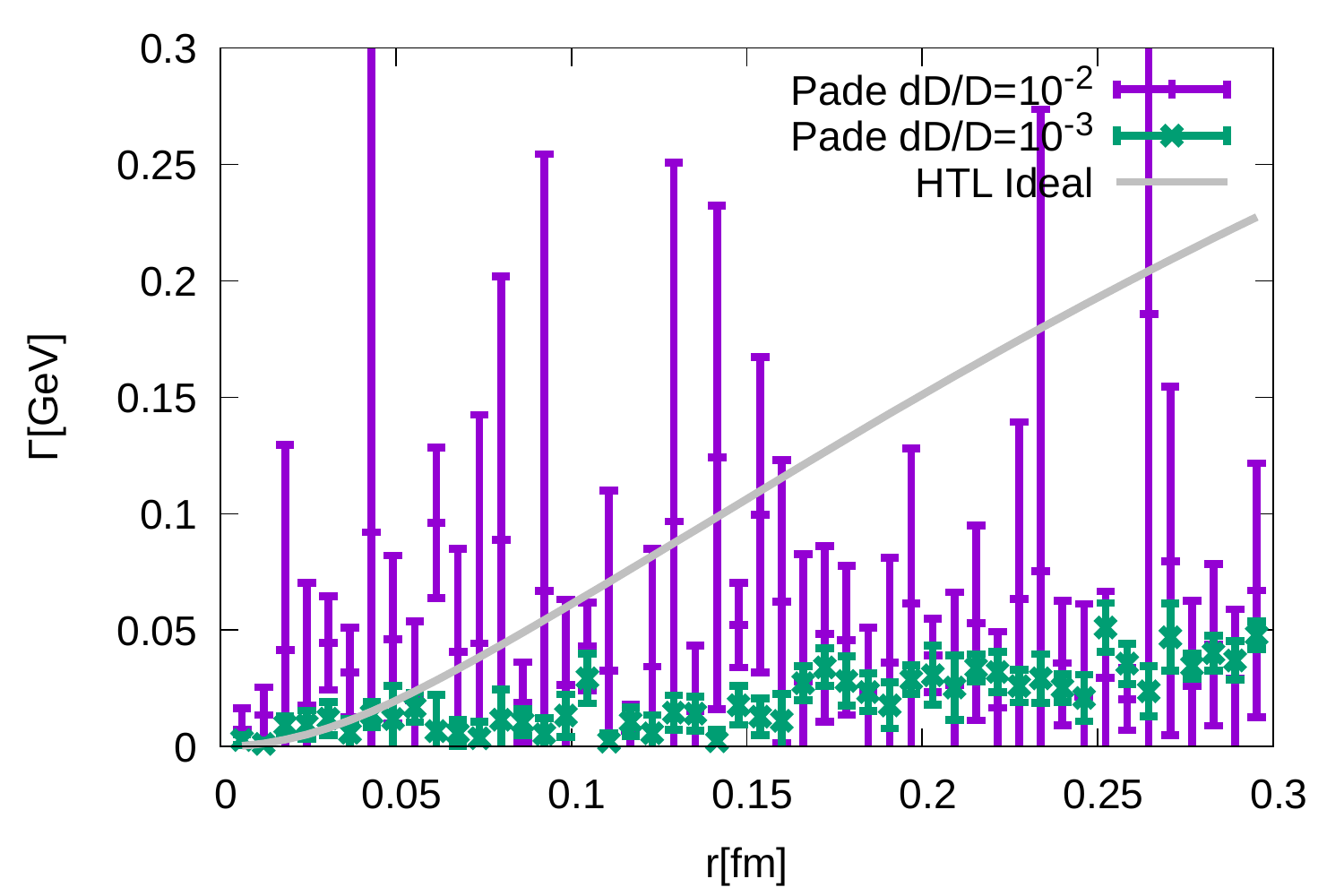}
\caption{Extraction of spectral position $\Omega$ and width $\Gamma$ of the dominant peak, based on Hard Thermal Loop mock data with $dD/D = 10^{-2}$ and $dD/D = 10^{-3}$ for $T=667$MeV using Pade. The error bars are obtained from Jackknive resampling.}
\label{Fig:HTL_PotPade}
\end{figure}

Pad\'e methods of rational approximation require usage of highly precise data. To test the reliability of this approach we use noisy Hard Thermal Loop (HTL) correlation functions. We compute HTL correlation functions for $T=667$MeV discretised to 12 lattice points. We then assign random noise of such that relative errors  $\Delta D/D=10^{-2}$ and $10^{-3}$ and generate 1000 samples each. We then  carry out Pade interpolation and pole analysis with results as shown in Fig \ref{Fig:HTL_PotPade}.

From the pole analysis of HTL mock data we see that we are able to recover the peak position $\Omega$ within uncertainties for relatively large errors  $\frac{\Delta D}{D}=10^{-2}$. The results for $\Delta D/D=10^{-3}$ are excellent. When we try to estimate the width of the peak the results are less encouraging with the true values being consistently underestimated for both  $\Delta D/D = 10^{-2}$ and $\Delta D/D = 10^{-3}$.

\begin{figure}
\centering
\includegraphics[scale=.5]{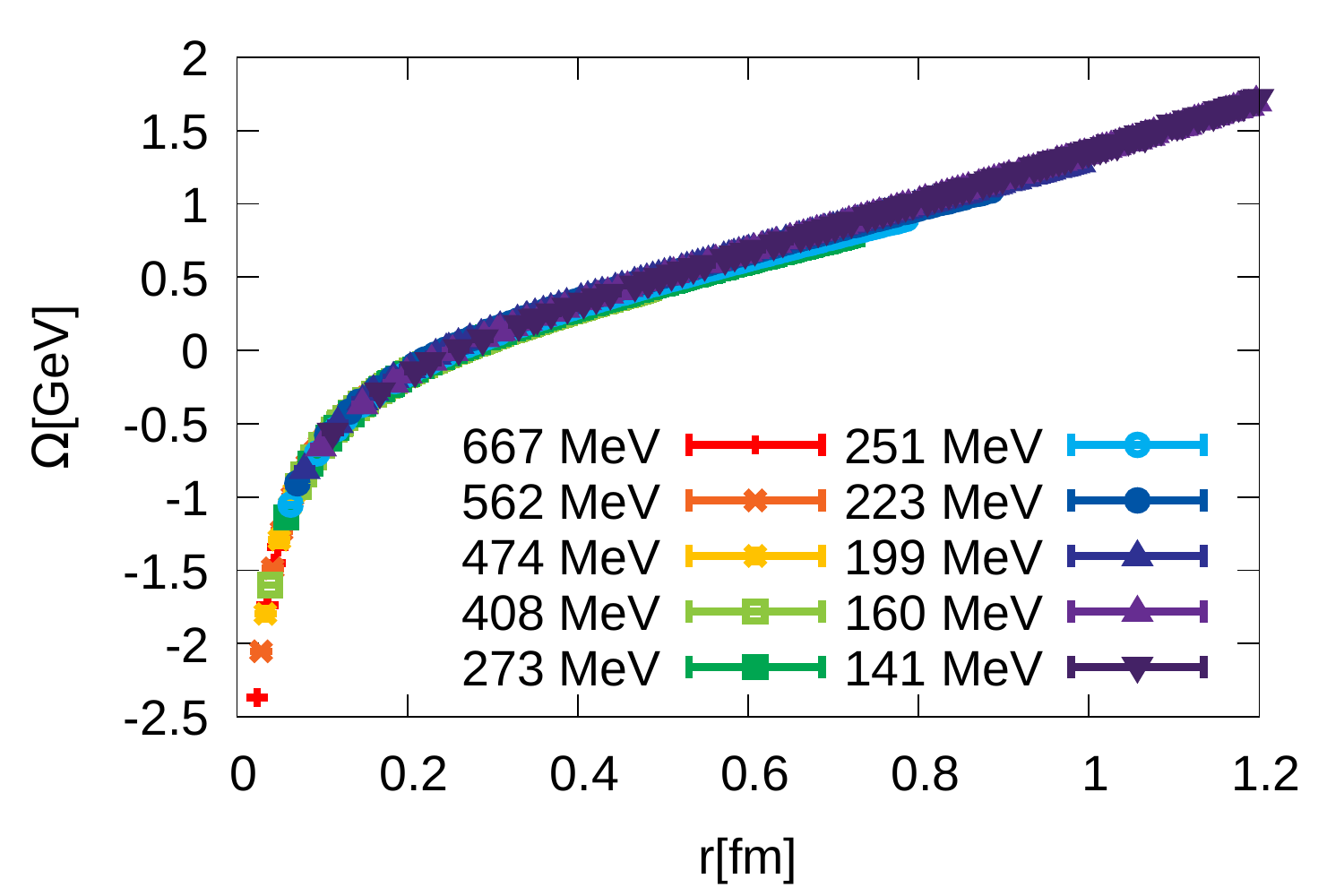}
\caption{$\Omega$ as a function of separation distance for different temperatures obtained from a Pad\'e pole analysis on $N_\tau = 12$. }
\label{Fig:Lattice_Pade}
\end{figure}

Given that the Pade is able to recover the peak position of the HTL mock data very well for error levels present in the actual data, we proceed to carry out the same analysis on the lattice correlators. Since the Pade underestimates the width we will show our analysis only for the peak position in Fig.\ref{Fig:Lattice_Pade}
Again the results obtained for $\Omega$ are very similar to the $T=0$ results and show no significant change with temperature, as was previously seen with the Gaussian fits.
\section{Spectral function extraction using Bayesian Method}
The Bayesian approach makes use of the Bayes theorem:
        \begin{equation}
            P[\rho|D,I]\propto P[D|\rho,I]P[\rho|I] = {\rm exp}[-L+\alpha S_{\rm BR}],
        \end{equation}
$P[\rho|D,I]$ is the posterior probability which is the probability of $\rho$ to be the correct spectrum given the data $D$ and prior information. $L$, the Likelihood, is the usual quadratic distance used in $\chi^2$ fitting. The prior probability acts as a regulator
\begin{equation}
    P(\rho |I) = \exp ( \alpha S_{BR}).
\end{equation}
Generally the choice of regulator depends on the particular choice of Bayesian method, however for this study we use the BR prior\cite{Burnier:2013nla}.
\begin{equation}
        S_{\rm BR}=\int d\omega \big( 1- \frac{\rho(\omega)}{m(\omega)} + {\rm log}\big[ \frac{\rho(\omega)}{m(\omega)} \big]\big),
\end{equation}
where $m(\omega)$ is the default model. We generally use the most uninformative default model m = const.

Unlike previous Bayesian approaches where the hyperparameter $\alpha$ is integrated out, here in the hope of removing ringing artifacts we draw inspiration from the Morozov Criteria and tune $\alpha$ such that $L = N_\tau/2$. Then we look for the most probable spectral function by locating the extremum of the posterior.

As with the Pad\'e, we test the reliability of Bayesian reconstruction with mock HTL euclidean data with errors as described in section \ref{pade}.
\begin{figure}
   \includegraphics[scale=0.5]{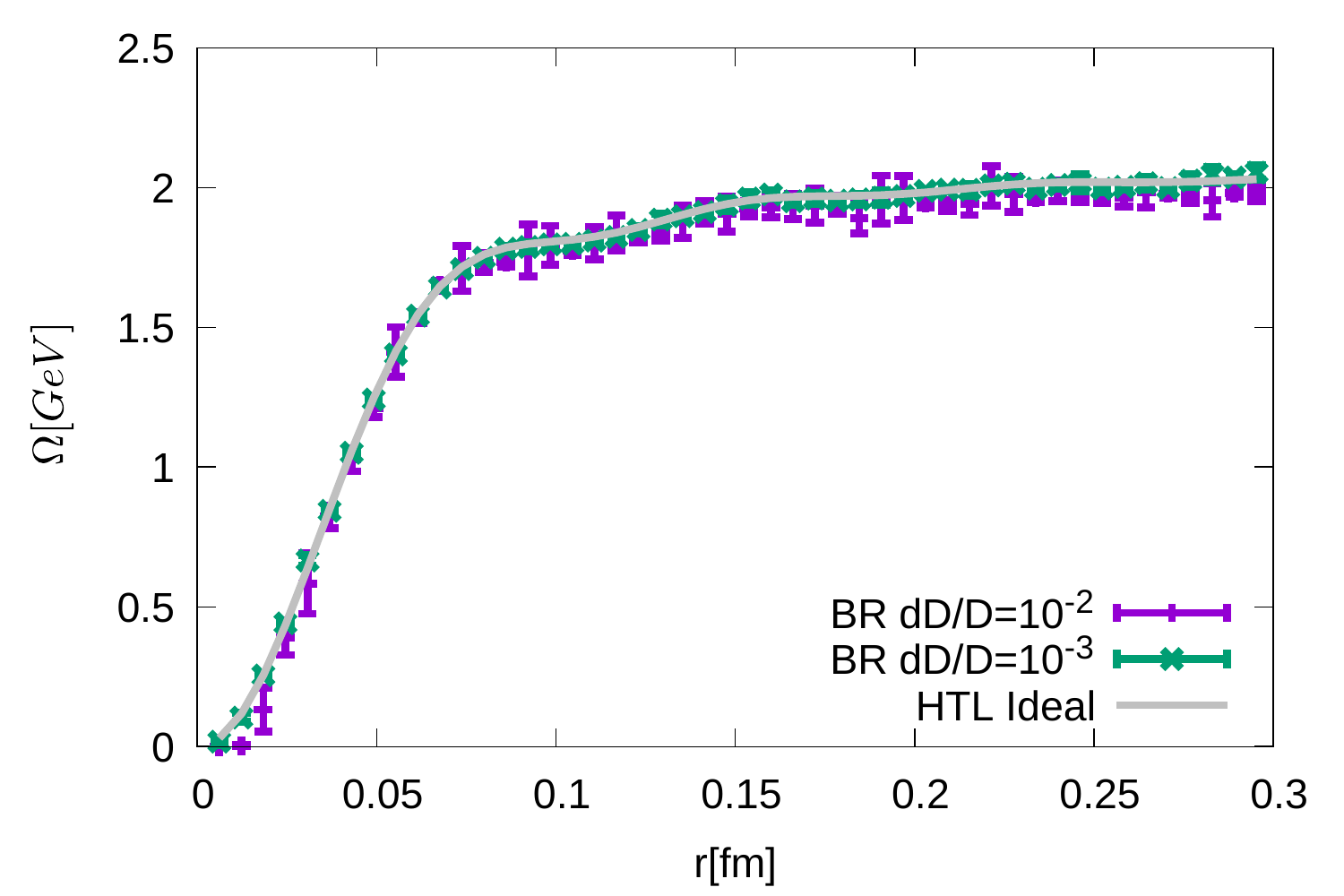}
    \includegraphics[scale=0.5]{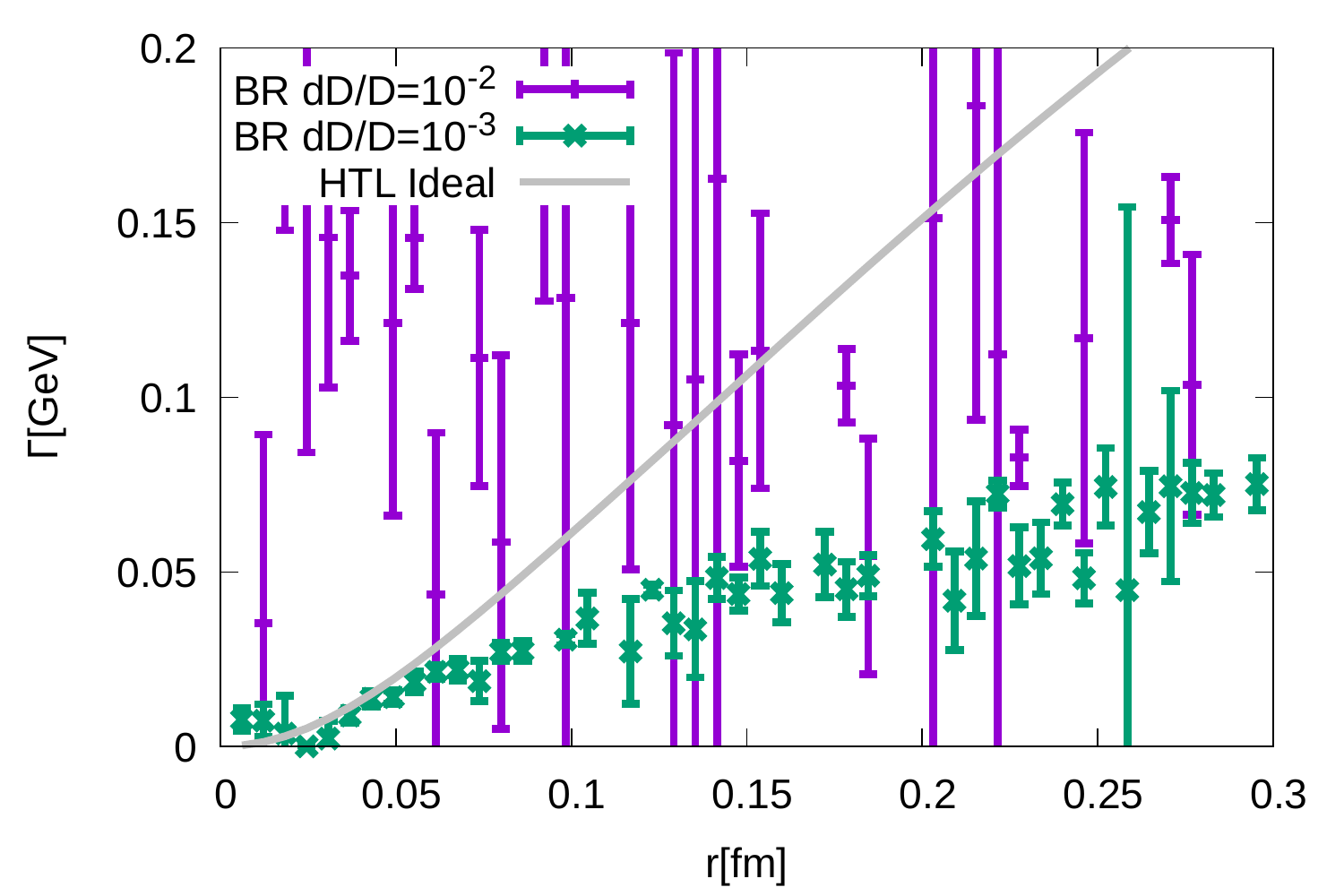}
\caption{Extraction of $\Omega$ and width $\Gamma$ for Hard Thermal Loop ideal data for $dD/D = 10^{-2}$ and $dD/D = 10^{-3}$ for $T=667$MeV using the BR method. The error bars are obtained from Jackknife resampling.}
    \label{br_pot}
   \end{figure}
Results for $\Omega$ and $\Gamma$ using Bayesian approach are shown in Fig.\ref{br_pot}. We see that the location of peak for $\Delta D/D = 10^{-2}$ and $\Delta D/D = 10^{-3}$ are very well estimated within uncertainties. The benchmark results show that the BR method is able to reproduce the peak position better than the Pad\'e for $\Delta D/D = 10^{-2}$. However, the width $\Gamma$ is still not reliable even for $\Delta D/D = 10^{-3}$. It does seem to be closer to the analytical result than the Pad\'e for $\Delta D/D = 10^{-3}$.

Even though we have successfully carried out the analysis on mock HTL data, we are challenged to doing the same on our lattice data.
Only at low temperatures (low $\beta$) the reconstruction converges. At high temperatures (high $\beta$) we observe that the effective mass plots show non-monotonous behavior at small distances. Fine lattices with improved gauge actions exhibit such effects at small time and distances  \cite{Bazavov:2019qoo}.
%The culprit here is the lattice spacing and this was first observed in \cite{Bazavov:2019qoo}.

\begin{figure}
\centering
\includegraphics[scale=.5]{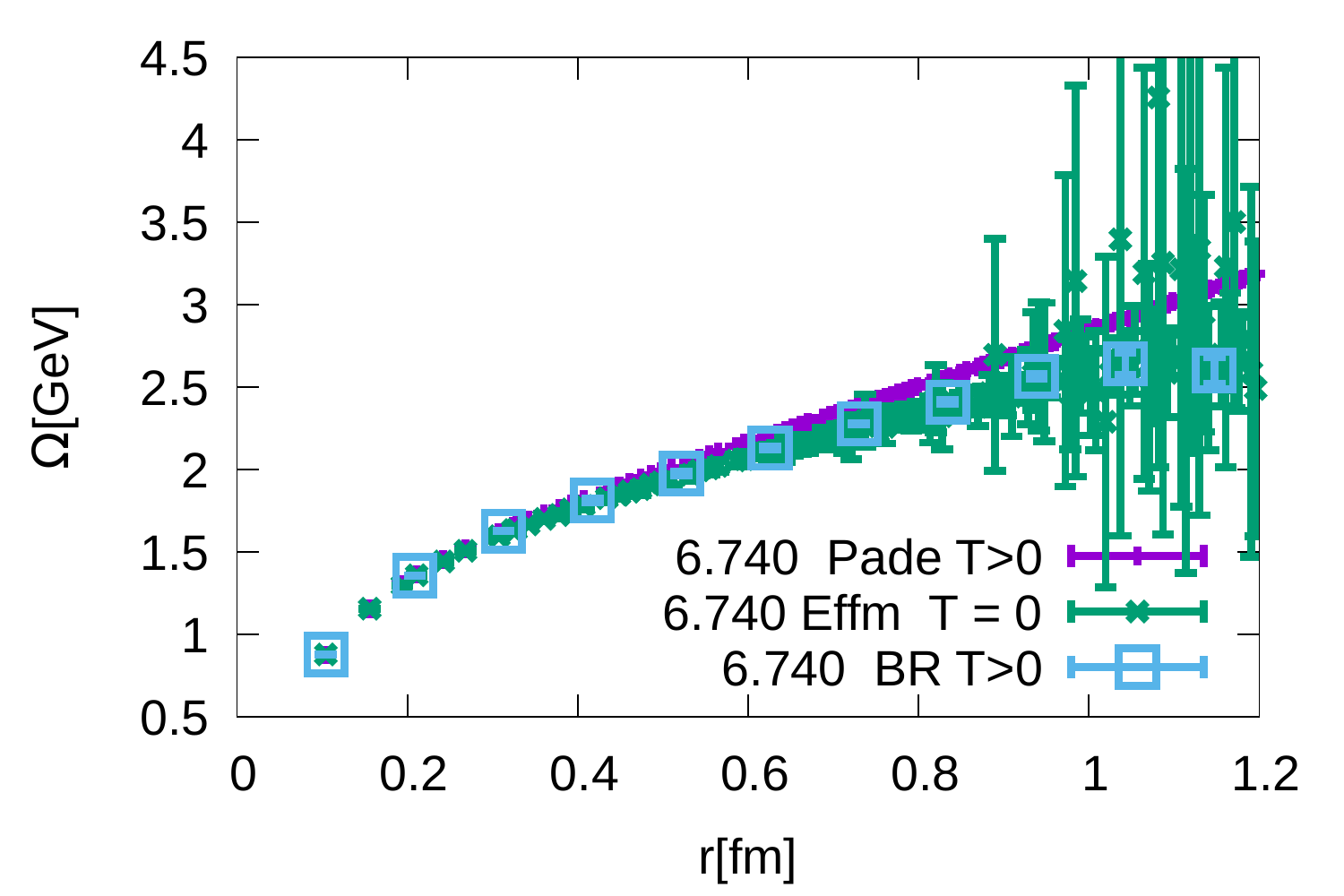}
\caption{Comparison of $\Omega$ using the Pad\'e and BR method at $\beta = 6.740$ with $N_\tau =12$ ($T=151$ MeV). The $T=0$ potential for the same $\beta$ are given as green data points.}
\label{br_lattice}
\end{figure}

This is a manifestation of positivity violation in the spectral function. Since the Bayesian is based on positive definite spectral function it cannot be applied for spectral reconstruction at high temperatures on our data. However, we can extract the peak position of the spectral function at low temperature (T=151 MeV) as shown in Fig.\ref{br_lattice}.

\section{Conclusion}
Here we presented a first study of the Wilson loop/line spectral structure on high precision HISQ ensembles. The peak position of the dominant spectral peak $\Omega$ obtained from Gaussian fits and Pade differs from some previous studies \cite{Burnier:2015tda} and \cite{Burnier:2014ssa} and our results do not show any significant modification with increasing temperature. 
However, are similar to an earlier analysis \cite{Petreczky:2017aiz} on a subset of the same data.

$\Gamma/T$ shows a linear increase with temperature in Gaussian fits, we have not shown the $\Gamma$ for Pad\'e since the results are systematically underestimated in our benchmarks. In the preprint \cite{Bala:2021fkm} and proceedings \cite{D.Bala:2021} we have also presented a different approach inspired by hard thermal loop perturbation theory \cite{Bala:2019cqu}.

\section*{Acknowledgements}

This material is based upon work supported by the U.S. Department of Energy, Office of Science, Office of Nuclear Physics through the (i) Contract No. DE-SC0012704, and  (ii) Scientific Discovery through Advance Computing (SciDAC) award  Computing the Properties of Matter with Leadership Computing Resources. (iii) R.L., G.P. and A.R. acknowledge funding by the Research Council of Norway under the FRIPRO Young Research Talent grant 286883. (iv) J.H.W.’s research was funded by the Deutsche Forschungsgemeinschaft (DFG, German Research Foundation) - Projektnummer 417533893/GRK2575 ``Rethinking Quantum Field Theory''. (v) D.B. and O.K. acknowledge support by the Deutsche Forschungsgemeinschaft (DFG, German Research Foundation) through the CRC-TR 211 'Strong-interaction matter under extreme conditions'– project number 315477589 – TRR 211.

This research used awards of computer time provided by: (i) The INCITE and ALCC programs at Oak Ridge Leadership Computing Facility, a DOE Office of Science User Facility operated under Contract No. DE-AC05- 00OR22725. (ii) The National Energy Research Scientific Computing Center (NERSC), a U.S. Department of Energy Office of Science User Facility located at Lawrence Berkeley National Laboratory, operated under Contract No. DE-AC02- 05CH11231. (iii) The PRACE award on JUWELS at GCS@FZJ, Germany. (iv) The facilities of the USQCD Collaboration, which are funded by the Office of Science of the U.S. Department of Energy. (v) The UNINETT Sigma2 - the National Infrastructure for High Performance Computing and Data Storage in Norway under project NN9578K-QCDrtX "Real-time dynamics of nuclear matter under extreme conditions".

\end{document}